\documentclass[a4paper,11pt]{article}
\pdfoutput=1 

\usepackage{jcappub} 
\usepackage{graphicx,color}
\usepackage{enumitem}
\usepackage{braket}
\usepackage{slashed}
\usepackage[utf8]{inputenc}
\usepackage{hyperref}
\usepackage{float}
\usepackage{caption}
\usepackage{subfigure}
\usepackage{makecell}
\usepackage{epsfig}
\usepackage{amsmath,amssymb}
\usepackage{hyperref}
\usepackage{ulem}

\hypersetup{
	colorlinks=true,
	citecolor=black,
	filecolor=black,
	linkcolor=blue,
	urlcolor=blue
}

\usepackage{dsfont}

\DeclareMathOperator{\Tr}{Tr}

\DeclareMathOperator{\der}{\mathrm{d}}
\DeclareMathOperator{\dif}{\, \text{d} \hspace*{-0.065cm}}
\DeclareMathOperator{\heta}{\eta}
\DeclareMathOperator{\hk}{\hat{\mathit{k}}}

\newcommand{\dof}[1]{g_{*,\tiny{#1}}}
\newcommand{\tem}[2]{T_{\tiny{#1}}^{(#2)}}
\newcommand{\lr}[1]{\left(#1\right)}
\newcommand{\tti}[1]{\text{\tiny{#1}}}

\usepackage{todonotes}

\title{Testing BSM Physics with  Gravitational Waves}

\author[a]{F. Muia,}
\author[a]{F. Quevedo,}
\author[b]{A. Schachner,}
\author[a]{G. Villa}

\affiliation[a]{\footnotesize DAMTP,  Centre for Mathematical Sciences,  University of Cambridge, Wilberforce Road,  Cambridge, CB3 0WA, UK}
\affiliation[b]{\footnotesize Department of Physics, Cornell University, Ithaca, NY 14853, USA}

\emailAdd{fm538@cam.ac.uk}
\emailAdd{fq201@damtp.cam.ac.uk}
\emailAdd{as3475@cornell.edu}
\emailAdd{gv297@cam.ac.uk}

\abstract{

The Cosmic Gravitational Wave Background (CGWB) is an irreducible background of gravitational waves generated by particle exchange in the early Universe plasma. Standard Model particles  contribute to such a stochastic background with a peak at $f\sim 80$ GHz. Any physics beyond the Standard Model (BSM) may modify the CGWB spectrum, making it a potential testing ground for BSM  physics.
We consider the impact of general BSM scenarios on the CGWB, including an arbitrary number of hidden sectors.
We find that the largest amplitude of the CGWB comes from  the  sector that dominates  the energy density  after reheating and confirm the dominance of the SM  for standard cosmological histories. 
For non-standard cosmological histories, such as those with a stiff equation of state $\omega >1/3$, like in kination, BSM physics may dominate and modify the spectrum substantially. 
We conclude that, if the CGWB is detected at lower frequencies and amplitudes compared to that of the SM, it will hint at extra massive degrees of freedom or hidden sectors.
If it is instead measured at higher values, it will imply a period with  $\omega >1/3$.
We argue that for scenarios with periods of kination in the early Universe, a significant fraction of the parameter space can be ruled out from dark radiation bounds at  BBN.
}

\begin{document} 
\maketitle
\flushbottom

\section{Introduction}
\label{sec:Intro}

Testing physics beyond the Standard Model (BSM) of particle physics is extremely challenging, especially for energies much higher than the TeV scale. Proton decay, tensor modes from the Cosmic Microwave Background (CMB) and possibly cosmic strings are  some of the very few potential signatures that can be searched in order to explore very high energies, like those of the Grand Unification scale, which are beyond reach of collider experiments, even in the long future. Gravitational waves, especially at ultra high frequencies, may provide a new tool to explore physics beyond the Standard Model at very high energies, as well as the early stages of the universe and it is worth exploring this potential.\vspace{0.1cm}

The Cosmic Gravitational Wave Background (CGWB) is an irreducible background of Gravitational Waves (GWs), sourced by the interactions of the constituents of a plasma in thermal equilibrium. Its shape was computed in~\cite{Ghiglieri:2015nfa,Ghiglieri:2020mhm} to leading-log and full leading order, respectively, for the Standard Model (SM) particles and interactions. This approach was later generalised in~\cite{Ringwald:2020ist} to the spectrum of an arbitrary weakly coupled renormalisable gauge theory with couplings to scalars and fermions. In these papers, it was shown that the shape of the CGWB is determined mostly by gauge interactions at the hottest moment of the evolution of the plasma. In other words, the CGWB carries a snapshot of how the Universe looked like right after reheating. \vspace{0.1cm}

The shape of the CGWB is therefore a direct probe for physics much further than the last scattering surface, which bounds electromagnetic astronomy. Due to its characteristic frequency, around $80$ GHz, the CGWB is an example of an ultra high frequency source of GWs, which are attracting the attention of a growing community of both theorists and experimentalists~\cite{Aggarwal:2020olq}. It is becoming increasingly clear that early Universe processes release backgrounds at these high frequencies, and that their potential detection provides a window for the direct study of high energy physics that is not achievable through electromagnetic astronomy or collider physics. \vspace{0.1cm}

In this paper, we study the general implications of physics beyond the Standard Model (BSM) on the spectrum of the CGWB. 
In particular, we study modifications of the CGWB due to the presence of general hidden sectors\footnote{
We focus on hidden -thermally independent- sectors 
defined as  particles coupled to the visible sector only through couplings with interaction rate satisfying $\Gamma \ll H$, so that they cannot reach equilibrium. Examples are gravitational and other non-renormalisable couplings to the SM  that generically appear in string theory.}.
In standard cosmology with inflation followed by radiation and matter domination,
we find that the sector dominating the energy density at reheating determines the peak of the CGWB spectrum.
However, despite the presence of hidden sectors,
it turns out that generically the SM dominates the spectrum of the CGWB, thereby generalising the findings of~\cite{Ringwald:2020ist}.  \vspace{0.1cm}

Further, we explore non-standard cosmologies and observe that, if there are periods where the background energy density is governed by a fluid with a stiff equation of state $\omega >1/3$ such as in kination domination,
then the contributions from BSM physics dominates. We show that the shape of the CGWB can carry information about the interactions among hidden sectors, and we manage to rule out part of a parameter space described in terms of the number of e-folds of kination and maximum temperature.  \vspace{0.1cm}

\begin{figure}[t!]
\centerline{\includegraphics[width=0.7\textwidth]{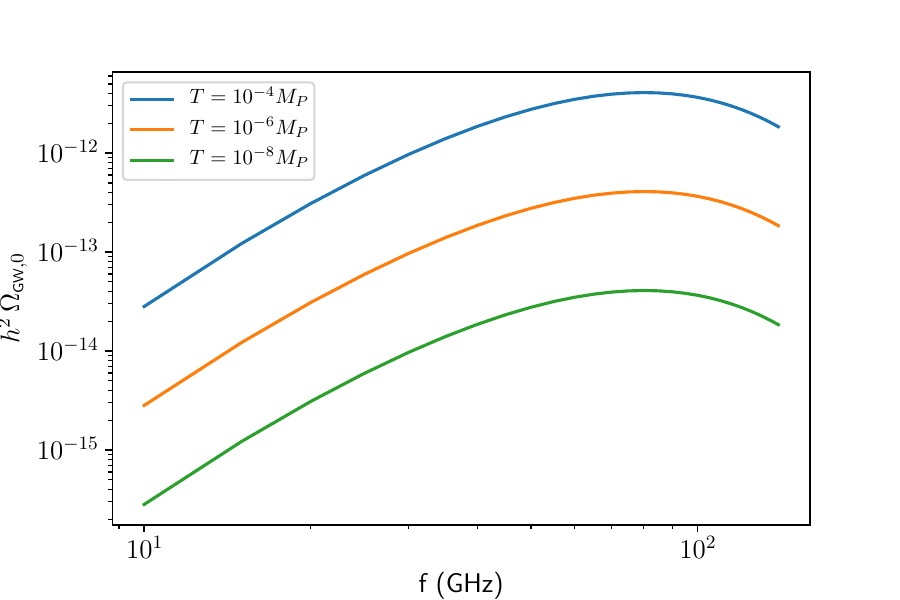}}
\caption{GW spectrum for the SM with different reheating temperatures and values $g_{h}=0.1$, $g_{w}=0.3$ and $g_{s}=0.5$ for hypercharge, weak and strong coupling constants, respectively.
We use $h^2\Omega_\tti{GW,0}$ with $h\sim 70$ for the present GW energy density to be independent of cosmological uncertainties.}
\label{fig:SMcase}
\end{figure}

More concretely, we focus on GWs sourced by a thermal bath in the early Universe due to particle collisions.
The amount of produced GWs is characterised by a fraction of the total energy density $\Omega_{\text{\tiny{GW}}, 0}(f)$.\footnote{We use the subscript $``0"$ to indicate that quantities are evaluated in the present Universe.}
When considering only the SM, the GW spectrum $\Omega_{\text{\tiny{GW}}, 0}(f)$ features a peak frequency around $80 \, $ GHz \footnote{Since the frequency tracks the CMB case and falls in the microwave range reference~\cite{Ringwald:2020ist} called this radiation CGMB. We prefer to slightly deviate from this notation since we will explore other frequency ranges and chose the more general CGWB. } and an amplitude which is linear in the reheating temperature $T_{\text{max}}$, corresponding to the maximum temperature the thermal bath has experienced.
In Fig.~\ref{fig:SMcase}, we report\footnote{We do not consider additional processes becoming relevant at very high temperatures~\cite{Ghiglieri:2022rfp}.} the GW spectrum obtained with $T_{\text{max}}=10^{-4}M_{P}, \, T_{\text{max}}=10^{-6}M_{P}$ and $T_{\text{max}}=10^{-8}M_{P}$, assuming that the entire particle content is attributed to the SM.
At the peak frequency, the energy density $\Omega_{\text{\tiny{GW}},0} (f_{\rm peak}) \sim 10^{-11}$ is well below the BBN bound $\Omega_{\text{\tiny{GW}},0}^{\text{\tiny{BBN}}} \sim 10^{-6}$.
It has already been shown in~\cite{Ringwald:2020ist} that, for a given $T_{\text{max}}$, it is challenging to enhance the amplitude of such a peak, at least in weakly coupled extensions of the SM with a standard cosmological history: despite the enhanced GW emission from additional interactions, the spectrum also experiences more red-shifting due to the emission happening earlier.
Indeed, since $H^2\sim g_{\tiny{*}} T^4$, for fixed $T$ the Hubble scale of emission is minimised by the SM because any hypothetical extension of the SM involves additional degrees of freedom. 

\subsection{Executive summary}\label{sec:ExecutiveSummary}

This work investigates the general potential for the CGWB to explore physics beyond the Standard Model of particle physics. 
We study general extensions of the SM including an arbitrary number of hidden sectors. In this vein, we write the energy density $\rho$ as $\rho = \sum_i \rho^{(i)}$ where $\rho^{(i)}$ corresponds to the energy density of the \textit{i}$^\text{th}$ sector.
In principle, each sector can have a different temperature $T_i$ as long as the constraint on dark radiation at BBN is satisfied.
The thermal bath in each decoupled sector produces a different GW background with energy density $\rho^{(i)}_{\text{\tiny{GW}}}$.
The position of the peak in each spectrum is determined by the temperature of the corresponding sector. Close to this peak, the time-dependent part of $\rho^{(i)}_{\tti{GW}}$ produced per e-fold in the sector $i$ at time $t$ is approximated by
\begin{equation}\label{eq:fraction}
    \frac{d}{d\log a}\lr{\frac{d \rho^{(i)}_{\tti{GW}}}{d \log k}} \sim T_i \; \frac{\rho^{(i)}a(t)^4}{\sqrt{\dof{tot}}}F(\hk)\,.
\end{equation}
Here, $a$ is the scale factor, $k$ the physical wavenumber of the GW, $g_{*, \tiny{tot}}$ the effective number of massless degrees of freedom and $F(\hat{k}_i)$ a function of $\hat{k}_i\equiv k/T_i$ which encodes the interactions that source GWs.

\begin{figure}[t!]
\centering
\includegraphics[width=.7\linewidth]{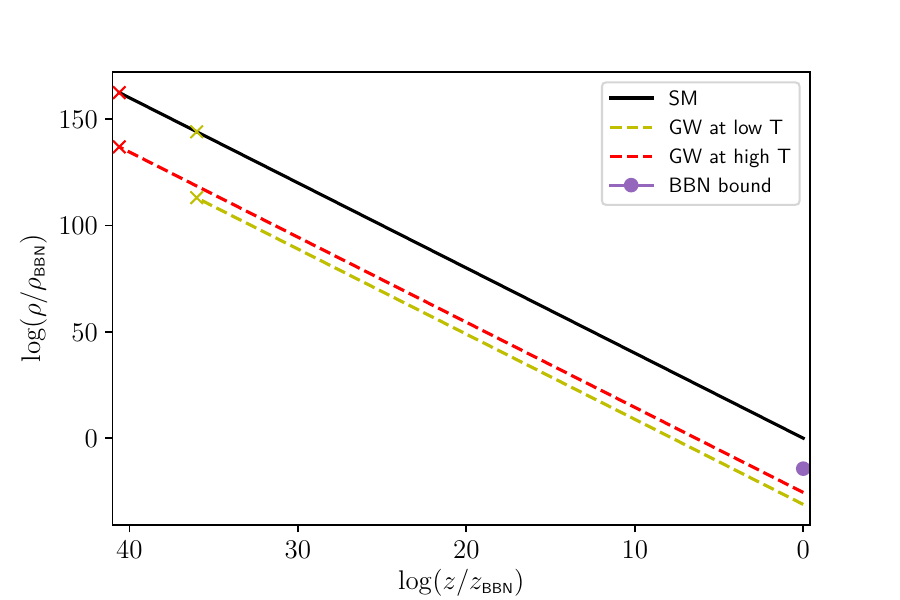}  
\caption{
Evolution of the energy density of the SM (black) and that of GWs (red, green) emitted at two different times (indicated by crosses), as a function of the redshift.
The origin is chosen so that the first emission occurs at $T_{\tti{max}}=10^{-4}M_P.$
As explained below Eq.~\eqref{eq:fraction}, and illustrated by the red and yellow line, GWs sourced later, when the bath is colder, are sub-dominant. 
}
\label{fig:scenario1}
\end{figure}

There are two main lessons to learn from Eq.~\eqref{eq:fraction}:
\begin{itemize}
    \item Assuming  that $g_{*, \tiny{tot}}=\text{const.}$ 
    Then, because $\rho^{(i)}\sim a(t)^{-4}$ and both $k$ and $T$ redshift as $1/a$, the only non-constant term in the right hand side is $T_i$. 
    Therefore, the amount of GWs released is larger at early times, when the sector was hotter, and these contributions dominate the GW energy density, as illustrated in Fig.~\ref{fig:scenario1}. 
    In other words, the CGWB carries an imprint of the Universe during its hottest phase and can be used as a thermometer of the hot big bang~\cite{Ringwald:2020ist}.
    \item Once hidden sectors are taken into account, several spectra of the schematic type shown in Fig.~\ref{fig:SMcase} for the SM overlie each other, thereby potentially resulting in novel shapes. 
    From Eq.~\eqref{eq:fraction}, we note that the dominant contribution to GWs at every time will be sourced by the sector dominating the energy density at that time. 
\end{itemize}
Putting both arguments together, because for every sector the amplitude of the CGWB is dominated at early times, the shape of the CGWB will be dictated by the sector carrying the most energy density after reheating.

\begin{figure}[t!]
\centering
\includegraphics[width=.7\linewidth]{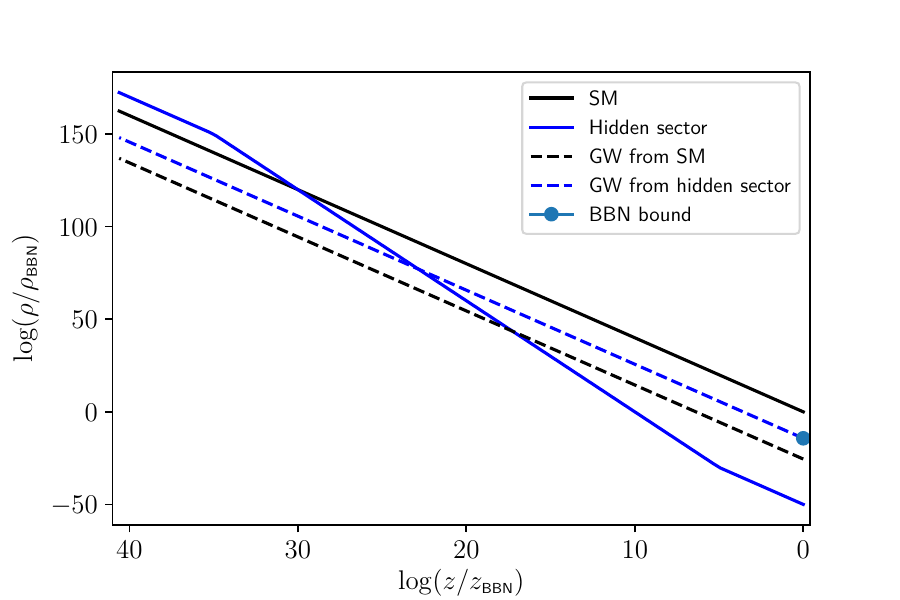}
\caption{Evolution of the energy densities in the kination scenario as a function of redshift. Solid lines represent massless degrees of freedom, and dashed lines the respective dominant contribution to $\Omega^{(i)}_{\tti{GW},0}$. The figure illustrates a hidden sector with a variable equation of state. During the period $\log(z/z_{\tti{BBN}})\in (5,35)$, the hidden sector undergoes kination, $\rho(t) \sim a(t)^{-6}$. For a sufficiently long kination era, GWs from the hidden sector can saturate the BBN bound while the energy content in massless degrees of freedom avoids it.}
\label{fig:scenario3}
\end{figure}

Throughout the paper, we study how this shape is modified in the presence of a hot hidden sector in the early Universe.
In order for such a scenario to be compatible with the BBN bound, the hidden sector energy density $\rho_h$ must satisfy
\begin{equation}\label{eq:BBNBound} 
\text{BBN bound: }\quad\rho_h \lesssim  \frac{\pi^2}{30}\times \frac{7}{4}\Delta N_{\text{\tiny{eff}}} T_{\rm vis}^4 \,,
\end{equation}
when the temperature of the visible sector is $T_{\rm vis}\sim 1 $ MeV. 
The latest Planck results~\cite{Planck:2018vyg} constrain $\Delta N_{\text{\tiny{eff}}} \leq 0.30$ at $95\%$ confidence level, which is a tight constraint on the existence of hot hidden sectors at BBN. 
A way to reconcile an early hot sector meeting Eq.~\eqref{eq:BBNBound} is through fast dilution of the energy density. 
This happens for instance if the relevant sector undergoes a phase with a stiff equation of state $\omega >1/3$, such as a period of kination, illustrated in Fig.~\ref{fig:scenario3}.
We will study this example more thoroughly in Sec.~\ref{sec:kination}.

To summarise, our main goal is a model independent study of the CGWB, establishing the robustness of the SM spectrum, even in the presence of general BSM physics, and exploring variations from it. We consider alternative cosmological histories  and perform a detailed analysis of such largely unexplored scenarios  by putting constraints on their parameters, such as reheating temperatures or the duration of kination eras. We find regions of parameter space saturating the BBN bound by obtaining $\Omega_{\text{\tiny{GW}}, 0} \sim 10^{-6}$ at the peak of the spectrum and regions which are already ruled out by the BBN bound.

\subsection*{Plan for the paper}

This paper is organised as follows.
In Sec.~\ref{sec:CGWBThermalPlasma},
we review the production of GWs from weakly coupled thermal plasmas, and their subsequent propagation through spacetime. 
We emphasise the effects of hidden sectors, and compute the shape of the CGWB sourced by each sector. 
For each of these spectra, we analyse the behaviour of the amplitude and peak frequency, thereby concluding that the behaviour of the total amplitude is controlled by the sector dominating the energy density after reheating.
We also show how our framework reproduces the \textit{standard scenario} considered in~\cite{Ghiglieri:2015nfa,Ghiglieri:2020mhm,Ringwald:2020ist}, in which the energy density of the Universe is dominated by the SM (or a minimal extension thereof) and the equation of state of the background is that of radiation.
In Sec.~\ref{sec:results},
we illustrate two ways in which a hidden sector may have dominated the energy density of the early Universe while respecting the BBN bound in Eq.~\eqref{eq:BBNBound}. These processes consist of injection of entropy through annihilation of massive degrees of freedom, or a stiff equation of state, and we will refer to them as the \textit{entropy scenario} and \textit{kination scenario}, respectively. We argue that the latter scenario can yield substantially different results, including a boost in the amplitude.
We summarise our results in Sec.~\ref{sec:conclusions}.

\section{CGWB from BSM Theories}\label{sec:CGWBThermalPlasma}

\subsection{CGWB from a thermal plasma}

GWs travel essentially freely after emission and, thus, carry unaltered information about processes that sourced them. 
For this reason, the CGWB constitutes a valuable target for directly observing pre-CMB physics. 
This section concerns the propagation of GWs from their time of emission to today. 
Specifically, we derive an expression for the fractional energy density $\Omega_{\text{\tiny{GW},0}}(f)$ in terms of the frequency today.\\

The GW energy density $\rho_{\text{\tiny GW}}^{(i)}$ in the $i^{\text{th}}$ sector is governed by the equation
\begin{equation}
\label{eq:cosmoprod}
 \dfrac{1}{a^{4}}\frac{d}{d t} \left(a^4 \rho_{\text{\tiny GW}}^{(i)}\right)=   (\partial_t+4H(t))\rho_\text{\tiny{GW}}^{(i)}=\frac{4T_i^4}{M_{P}^2}\int{\frac{\der^3 \textbf{k}}{(2\pi) ^3} \heta_i (T_i, \hk_i)} \,,
\end{equation}
where $a(t)$ is the scale factor, $H(t) = \dot{a}(t)/a(t)$ is the Hubble parameter, $T_i$ is the temperature of the $i^{\text{th}}$ sector and the right hand side plays the role of a source term.\footnote{We assume that the independent radiation sectors are either in thermal equilibrium or completely decoupled (thus not sourcing GWs). The only out of equilibrium process relevant for our purposes is that of fields becoming massive (and eventually annihilated), which can be considered adiabatic if $\Gamma \gg H$~\cite{Bernstein:1988bw}, with $\Gamma$ the effective interaction rate with the plasma.} 
The term $\eta_i(T_i, \hat{k}_i)$ encodes the effects of GW emission from a thermal plasma. 
While it is the sum of various components (see Appendix~\ref{sec: integrate omega} for the details), in our scenarios the dominant contribution comes from the Hard Thermal Loop (HTL) piece\footnote{We work sufficiently close to the peak of this function, $\hk \sim 4$, where the description in terms of quasiparticles is valid~\cite{Ghiglieri:2015nfa,Ghiglieri:2020mhm,Ringwald:2020ist}.} $\eta_{\text{\tiny{HTL}}}$ given by
\begin{equation}\label{eq:eta_HTL}
    \heta_\text{\tiny{HTL}}=  \sum_n\mathcal{N}_n\hat{m}_n^2(T)\xi_n(\hk) \,,\qquad \xi_n(\hk)=\frac{\hk}{16\pi (e^{\hk}-1)}  \log \left ( 1+\frac{4\hk^2}{\hat{m}_n^2} \right ) \,.
\end{equation}
Here we consider a theory with $n=1,\ldots,N_g$ gauge groups, gauge couplings $g_n(T)$, $\mathcal{N}_n$-dimensional Lie algebras and thermal masses $m_n(T)^2 = T^2 \, \hat{m}_n(T)^2 \propto T^2 \, g_n(T)$, whose expressions are given explicitly in Appendix~\ref{sec: integrate omega}. 
Also, we work under the assumption that $\rho_\text{\tiny GW}$ is always a sub-leading component of the energy density of the Universe.

We can find the total energy density in GWs today $\rho_{\text{\tiny GW},0}$ by integrating Eq.~\eqref{eq:cosmoprod}.
Assuming that the production of GWs starts at time $t_p$ and finishes at $t_{\rm end}$, we obtain
\begin{equation}\label{eq:rhoia}
\rho_{\text{\tiny GW}}^{(i)}(t_{\rm end}) = -\frac{2\sqrt{3}}{\pi^2M_{P}}\, \dfrac{1}{a^4_{\rm end}}\int_{t_{\rm end}}^{t_p}{\dif a \/\ T_i^7 a(t)^3 \int d\log k \, \frac{\hk_i^3\heta_i(T_i,\hk_i)}{\sqrt{\rho}} }\,,
\end{equation}
where we defined $a_{\rm end} \equiv a(t_{\rm end})$, $a_p \equiv a(t_p)$ and $a_0 \equiv a(t_0)$.
The total energy density of the Universe $\rho$ reads in terms of relativistic\footnote{Note that GW production occurs before BBN, where we assume that the Universe is radiation dominated.} degrees of freedom $g_{*,j}$
\begin{equation}
\rho=\frac{\pi^2}{30}\sum_j g_{*,j}\, T_j^4\, .
\end{equation}
From Eq.~\eqref{eq:rhoia} one can easily find the energy density in GWs today (in the $i^{th}$ sector), red-shifting the result by a factor $(a_{\rm end}/a_0)^4$.

Stochastic GW backgrounds are characterised by the energy density fraction per logarithmic frequency interval~\cite{Caprini:2018mtu} which can be computed as
\begin{equation}\label{eq:SGWB0} 
h^2 \Omega^{(i)}_{\text{\tiny GW},0} = \frac{h^2}{\rho_c} \left(\frac{a_{\rm end}}{a_0}\right)^4 \left.\frac{d \rho^{(i)}_{\text{\tiny GW}}}{d \log k}\right|_{\rm end} \,,
\end{equation}
where $\rho_c$ is the critical density and $H_0 = 100 \, h \, \text{Km} \, \text{sec}^{-1} \, \text{Mpc}^{-1}$ is the Hubble constant today.
Lastly, combining Eq.~\eqref{eq:rhoia} and Eq.~\eqref{eq:SGWB0} as well as using the current radiation energy density~\cite{Fixsen:2009ug}
\begin{equation}
\Omega_{\gamma, 0}=\frac{\rho_{\gamma,0}}{\rho_{c}}=\frac{\pi^2}{15}\frac{T_{{\rm vis},0}^4}{\rho_{c}}=2.4728(21)\times 10^{-5}/h^2\, ,
\end{equation}
we arrive at our final expression
\begin{equation}\label{eq:ampl}
    \dfrac{ \Omega^{(i)}_{\text{\tiny{GW}},0}}{ \Omega_{\gamma,0}}=  \frac{\lambda}{M_{P}}  \frac{1}{(a_0 T_{{\rm vis},0})^4}
    \int_{t_p}^{t_{\rm end}}\dif a \, \frac{(aT_i)^5}{a^2}\, \dfrac{T_{i}^{2}}{\sqrt{\rho}} \, \hk_i^3 \heta_i (T_i, \hk_i ) \,,
\end{equation}
in terms of
\begin{equation}
\lambda=\dfrac{30 \sqrt{3}}{\pi^{4}}\, , \qquad \dfrac{\sqrt{\rho}}{T_{i}^{2}}=\sqrt{\dfrac{\pi^{2}}{30}\sum_j{g_{*,j}\left ( \frac{T_j}{T_{i}} \right )^4}}\, .
\end{equation}

This is the energy density of GWs emitted by each independent sector.
In the remainder of this paper,
we use Eq.~\eqref{eq:ampl} to engineer scenarios with phenomenologically appealing features for $\Omega_{\text{\tiny{GW},0}}(f)$ by synergising hidden sectors with non-conventional cosmological phases.
But before delving into the specifics,
let us extract universal properties from the above results.

\subsection{General features}\label{sec:generalresults}

\subsubsection{Spectrum}

We begin by exploring the implications of Eq.~\eqref{eq:ampl} on the GW spectrum.
To this end,
let us consider a phase in the history of the Universe in which the number of relativistic degrees of freedom $g_{*,i}$ in the \textit{i}$^{th}$ sector is constant. 
Although GWs can be emitted for varying $g_{*,j}$, such contributions come on top of the results reported here, which in this sense can be regarded as a lower bound on the production of GWs.
However, as we discuss in Appendix~\ref{sec: integrate omega}, it is this contribution that features the qualitative aspects of the spectrum.

Let us write $\tem{i}{r}\equiv T_i(t^{r})$, $a_{r}\equiv a(t^{r})$, where $t^r$ indicates the time of reheating. Then, the energy densities of GWs in the sector $i$ emmitted after reheating is given by
\begin{equation}\label{eq:OmegaSector}
\dfrac{\Omega^{(i)}_{\text{\tiny{GW}},0}}{ \Omega_{\gamma,0}}\simeq  \lambda\,  \left (\frac{a_{r}}{a_0}\frac{\tem{i}{r}}{\tem{\text{vis}}{0}}\right)^4
     \frac{\lr{\tem{i}{r}-T_i(t^f)}}{M_{P}} \dfrac{\left (\tem{i}{r}\right )^{2}}{\sqrt{\rho^{(r)}}}  \hk_i^3 \heta_i \left (\tem{i}{r}, \hk_i \right )  \,.
\end{equation}

This follows from the integration of Eq.~\eqref{eq:ampl} over a phase in which the effective degrees of freedom are constant, and ignoring the logarithmic dependence of the couplings on $T$,
i.e., by treating $\eta$ as approximately constant.
The time $t^f$ indicates the moment in which effective degrees of freedom cease to be constant, an example being some of the fields becoming massive. 
Notice, however, that in general $\tem{i}{r}\gg T_i(t^f)$ so we can safely neglect the latter.
The total energy density of GWs produced by the thermal plasma during this phase amounts to
\begin{align}\label{eq:mastf}
\dfrac{ \Omega_{\text{\tiny{GW}},0}}{\Omega_{\gamma,0} } \simeq \lambda\,
    \sum_{i=1}^{N}{ \frac{\tem{i}{r}}{M_{P}}  \left (\frac{\tem{i}{r}}{\tem{vis}{0}}\frac{a_{r}}{a_0}\right)^4
     \dfrac{\left (\tem{i}{r}\right )^{2}}{\sqrt{\rho^{(r)}}} \, \hk_i^3 \heta_i \left (\tem{i}{r},\hk_i \right )} \,,
\end{align}
where $N$ denotes the number of thermally independent sectors.\\
A few comments are in order.
For the case where there is only one sector (the SM or some extension at high temperatures) with fixed equation of state, the quantities $a_r\tem{i}{r}/(a_0\tem{vis}{0})$ and $k/T$ are fixed by the conservation of co-moving entropy, up to high energy degrees of freedom (see Sec.~\ref{sec:entropy}). 
In addition, $T^2/\sqrt\rho$ is temperature-independent, and so is $\eta$ by assumption.
The total contribution is therefore linear in ${T}/{M_{P}}$, which implies that the CGWB could be used as a thermometer of the Hot Big Bang~\cite{Ringwald:2020ist}. 

Furthermore, as anticipated in Sec.~\ref{sec:Intro}, it also follows that the dominant contribution is sourced at the time where the plasma was the hottest (hence the superscript $r$, denoting \textit{reheating}, since this moment should happen after reaching thermal equilibrium, right after reheating).
We henceforth focus in such regime, commenting further in additional contributions to the CGWB in Appendix~\ref{sec: integrate omega}.
The rest of the section is devoted to the study of Eq.~\eqref{eq:OmegaSector}.

\subsubsection{Peak amplitude}

We observe in Fig.~\ref{fig:SMcase} that the shape of the SM contribution to Eq.~\eqref{eq:mastf} has a global maximum. This is true universally for the GW spectrum produced in each sector,\footnote{At least in weakly coupled extensions. Similar results at strong coupling have been studied in~\cite{Castells-Tiestos:2022qgu}.} as we will show momentarily.
In this subsection,
we study how the position and maximum amplitude of this peak, which are the phenomenologically most interesting aspects of the spectrum, change within the different scenarios.
In order to obtain an analytic understanding, we restrict ourselves to consider the leading-log HTL part~\cite{Ghiglieri:2015nfa,Ghiglieri:2020mhm,Ringwald:2020ist}
of Eq.~\eqref{eq:eta_HTL}.\\

Let us define the objects
\begin{equation}
    F(\hk)=\lambda\, h^2\Omega_{\gamma,0}\hk^3\xi(\hk) \, ,\qquad K_i = \sum_{n(i)} \mathcal{N}_n\hat{m}^2_n(T) \,,
\end{equation}

where $\xi$, as defined in Eq.~\eqref{eq:eta_HTL}, depends only logarithmically on the field content of the scenario under consideration. Neglecting this dependence allows us to extract general features about the position of the peak and its amplitude.
Further, $K_{i}$ is a sum over all coupled gauge fields within the $i^{\text{th}}$ sector.
Then we can express the amplitude of an arbitrary coupled sector in terms of $F(\hk)$ as
\begin{equation}\label{eq: omega f}
    h^2\Omega^{(i)}_{\text{\tiny{GW}},0} \sim \left(\frac{a_{r}}{a_0} \frac{\tem{i}{r}}{\tem{vis}{0}}\right)^4 \frac{\tem{i}{r}}{M_{P}} \frac{\left(\tem{i}{r}\right)^2}{\sqrt{\rho^{(r)}}}\,  K_i\, F(\hk_i) \,.
\end{equation}
This expression formalises the schematic form of Eq.~\eqref{eq:fraction}. 
Indeed, $T_i^2/\sqrt{\rho}$ counts the degrees of freedom that dominate the energy density, and $K_iT_i^4$ is proportional to the energy density of the gauge fields in the $i^{\text{th}}$ sector, which source the HTL contribution.\\

The result in Eq.~\eqref{eq: omega f} will play a vital role in Sec.~\ref{sec:results} where we aim at boosting the spectrum in order to find larger signatures from GWs.
To this end,
we will consider cosmological periods with equations of state $\omega>1/3$, which increase the ratio $a_{r}\tem{i}{r}/(a_0\tem{vis}{0})$ appearing in Eq.~\eqref{eq: omega f}. Physically, these scenarios lead to larger energy densities than the standard scenario, thereby increasing the GW signal.

\subsubsection{Peak frequency}

The amplitude in Eq.~\eqref{eq: omega f} depends on the GW frequency $f$ through the quantity $\hk_i\equiv k/T_i$, which is evaluated at the time of production. To relate the latter to the frequency today, we use $k(t)a(t)=\rm const.$ to write
\begin{equation}\label{eq:freqPeak} 
    \hk_i=  \frac{2 \pi f}{T_i(t)} \frac{a_0}{a(t)} \quad \Rightarrow\quad \frac{f}{\tem{vis}{0}}\sim \frac{a_{r}}{a_0} \frac{\tem{i}{r}}{\tem{vis}{0}}\, .
\end{equation}
The position of the peak in Eq.~\eqref{eq: omega f} is determined by $F(\hk)$ at around $\hk \simeq 4.2$~\cite{Ringwald:2020ist}.\footnote{Small variations to this number arise from the logarithmic dependence of the HTL on the field content, and from additional leading order corrections.} 
Let us make contact with the standard scenario, where the ratio 
$a_{r}\tem{i}{r}/(a_0\tem{\text{vis}}{0})$ 
is fixed up to high energy degrees of freedom, namely
\begin{equation}\label{eq: freq}
    \frac{f}{\tem{\text{vis}}{0}}=\frac{\tem{i}{0}}{\tem{\text{vis}}{0}}\frac{\hk_i}{2\pi}\lr{\frac{\dof{i,0}}{\dof{i}}}^{1/3}\,,
\end{equation}
where the expression follows from conservation of co-moving entropy, which we explicitly show in Sec.~\ref{sec:entropy}. We have written $g_{*,i,0}$ as the current number of degrees of freedom in the $i^{\text{th}}$ sector, while $g_{*,i}$ is the number of degrees of freedom in the $i^{\text{th}}$ sector at the time of GW production.
The BBN bound requires that for any putative hidden sector the relation $\tem{i}{0}\leq\tem{\text{vis}}{0}$ holds, which implies that the maximum frequency is determined by the visible sector. Within one-sector BSM theories, the peak frequency is hence maximised by the SM since this gives the minimal contribution to $g_*$.\\

In the case of stiff equations of state, however, we can increase the ratio $ a_{r}\tem{i}{r}/(a_0\tem{\text{vis}}{0})$ and, by Eq.~\eqref{eq:freqPeak}, the peak frequency. An observation of the CGWB at a frequency bigger than the SM prediction would be a signal of a period with stiff equation of state. As we discuss in the next section, such scenarios typically yield a large amplitude.

In summary, the peak frequency is $f\sim T$ with a $\mathcal{O}(1)$ proportionality constant. Hence one can obtain the frequency at emission from the temperature of a given sector. In standard cosmology, at any time we know $T$ up to high energy degrees of freedom, so measurements of $T_0$ fix this frequency, which turns out to be $f \sim 80 \, \text{GHz}$. In the next sections we will see that an exotic equation of state allows for a larger $T$ than what is expected at a given redshift, thus allowing a larger frequency.

\subsubsection{The Standard Model}
\label{sec:VisibleSector}

In this subsection we reproduce the results of \cite{Ghiglieri:2020mhm} for the leading-order contribution to the CGWB predicted by the SM or an extension thereof belonging to the visible sector.

We thus consider a single sector and a single phase.
Hereby, we neglect the contributions to the CGWB after the EW phase transition because the amplitude is effectively linear with the maximum temperature, recall the discussion below Eq.~\eqref{eq:mastf}.\\

We can use Eq.~\eqref{eq: freq} to find from Eq.~\eqref{eq:mastf}
\begin{align}
\dfrac{\Omega_{\text{\tiny{GW}},0} }{ \Omega_{\gamma,0}}&=\lambda\,  
     \frac{\tem{\text{\tiny{SM}}}{\text{r}}-\tem{\text{\tiny{SM}}}{\text{\tiny{EW}}}}{M_{P}}
     \frac{\dof{\text{vis},0}^{1/3}}{\dof{\text{\tiny{SM}}}^{5/6}} \left(\frac{2\pi f}{\tem{\text{vis}}{0}}\right)^3
    \heta \left(\tem{\text{\tiny{SM}}}{\text{r}},\frac{2\pi f}{\tem{\text{vis}}{0}}\left(\frac{\dof{\text{\tiny{SM}}}}{\dof{\text{vis},0}}\right)^{1/3} \right) \,,
\end{align}
which is in perfect agreement with Eq.(2.21) of~\cite{Ringwald:2020ist} and depicted graphically in Fig.~\ref{fig:SMcase}.\\

We stress that the generalisation of the amplitude for weakly coupled extensions of the SM is obtained by simply replacing $g_{*,\text{\tiny{SM}}} \rightarrow g_{*,\text{\tiny{BSM}}}$ and $\eta_{\text{\tiny{SM}}} \rightarrow \eta_{\text{\tiny{BSM}}}$. The power of effective degrees of freedom arises from $\hat{k}^3$, $\rho^{(r)}$ and $ a_{r}\tem{i}{r}/(a_0\tem{\text{vis}}{0})$.
Adding new bosonic or fermionic particles only yields a small change since $\eta$ is dominated by gauge fields through the HTL contribution.
The modifications compared to the SM prediction are thus marginal for such BSM theories as already pointed out in~\cite{Ringwald:2020ist}.

\section{Novel CGWB scenarios}\label{sec:results}

The aforementioned results imply that the particle content and interactions of the sector dominating the energy density of the early Universe are observable through their gravitational imprint on the CGWB.
In this section we explore two scenarios involving a hidden sector that is hotter than the visible sector, whilst compatible with the BBN bound:
\begin{itemize}
\item the first scenario amounts to the injection of entropy in the visible sector through the annihilation of degrees of freedom.
\item the second scenario considers a stiff equation of state $P=\omega \rho$ with $\omega>1/3$.
\end{itemize}
The second scenario will actually turn out to be the most phenomenologically appealing as will be explained momentarily.

\begin{figure}[t!]
\centerline{\includegraphics[width=0.9\textwidth]{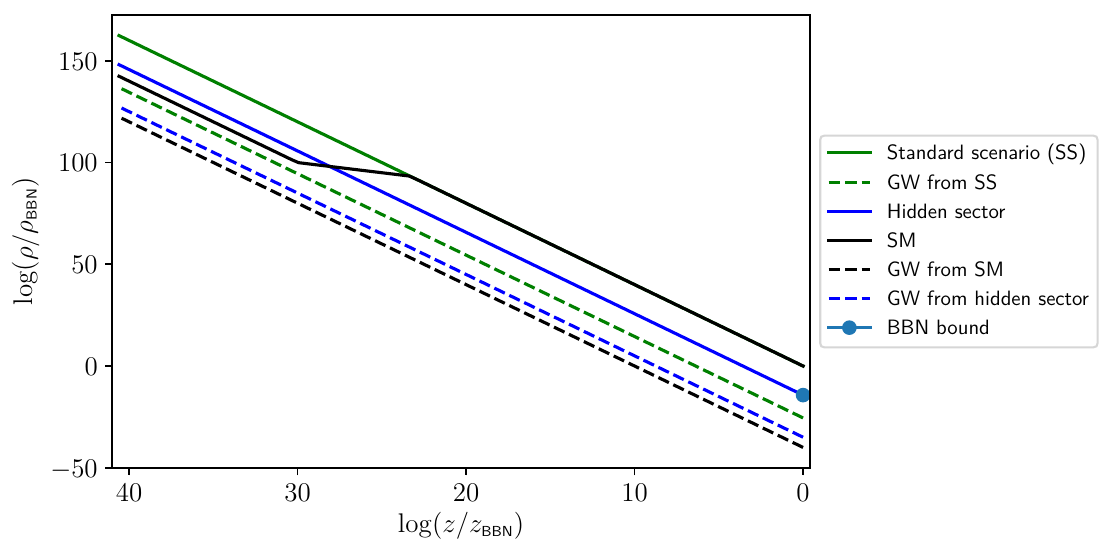}}
\caption{Energy density as a function of redshift in the entropy scenario, with SM (black)+hidden (blue) sector radiation (solid) and GW (dashed) energy densities. The standard scenario (green) is also plotted for comparison. For a given reheating temperature, the CGWB sourced in the standard scenario is always bigger.}
\label{fig:EntropyInjection}
\end{figure}

\subsection{Scenario 1 -- Entropy injection}\label{sec:entropy}

In this scenario, we make use of the following basic fact:
if there is some
field content in the early Universe that eventually gets massive whilst in thermal equilibrium, there is an injection of entropy into the remaining massless degrees of freedom. This leads to the temperature to redshift slower than $1/a$, the sector is being `heated up'. An example of this process is the well-known electron-positron annihilation, which renders the CMB hotter than the neutrino background. In this subsection we study how this heating up of the visible sector can induce different shapes of the CGWB.

\subsubsection*{Injection with one sector}

Initially, we consider a situation with only a visible sector as some extension of the SM.
The sector is effectively colder than in the standard scenario, as illustrated by the black solid line in Fig.~\ref{fig:EntropyInjection}. 
This results in a spectrum that peaks at smaller frequencies, according to Eq.~\eqref{eq: freq}, and a different (smaller) ratio $a_{r}\tem{\text{vis}}{r}/(a_0\tem{\text{vis}}{0})$. This is immediate to see from conservation of co-moving entropy $s(t)a(t)^3=\text{const.}$,
\begin{equation}\label{eq:comoving}
    \frac{a_r\tem{i}{r}}{a_0\tem{\text{vis}}{0}}=\lr{\frac{g_{*,\tti{0}}}{g_{*,\tti{r}}}}^{1/3} \, .
\end{equation}
The amplitude is therefore smaller than that of the SM, as already pointed out in~\cite{Ringwald:2020ist}.\\

An important shift in the frequency would have happened if the field content decreased a lot. 
From Eq.~\eqref{eq: freq}, we observe that, in order to redshift the frequency by an order of magnitude, one requires $g_{*,\text{\tiny{BSM}}}=10^3 g_{*,\text{\tiny{SM}}} \sim 10^5.$ 
This might have happened in multi-field inflation, for the inflatons reaching thermal equilibrium with the bath.
The diagram in Fig.~\ref{fig:EntropyInjection} and the fact that $h^2\Omega_{\tti{GW,0}}$ depends on the fourth power of the ratio~\eqref{eq:comoving} show the conclusions of~\cite{Ringwald:2020ist} that, with standard equation of state, the standard scenario shows the biggest amplitude for the GW spectrum. 
This remains true in the presence of hidden sectors, although there is the possibility of a richer phenomenology, as we now argue. 

\subsubsection*{Injection with two sectors}\label{sec: inject2sectors}

When the injection of entropy occurs as described above, it can happen that an initially cold sector becomes effectively hotter. We assume that this injection happened in the visible sector in order to avoid the BBN bound by ensuring its energy density is sufficiently large compared to that of the hidden sector. The scenario is illustrated by the black and blue lines in Fig.~\ref{fig:EntropyInjection}:
initially, the hidden sector is hotter, thus sourcing the largest contribution to the CGWB. However, due to injection of entropy, the redshift of the visible sector is slower and eventually dominates the energy density of the Universe.

Furthermore, there is the possibility that the spectrum features two peaks. This is because, as discussed before, the peak frequency of the visible sector is smaller. Thus, two peaks could appear in the particularly tuned case where the amplitude of the hidden sector and the BSM theory are similar, $\Omega^{(\text{\tiny{vis}})}_{\text{\tiny{GW}},0} \sim \Omega^{(\text{\tiny{h}})}_{\text{\tiny{GW}},0}$, and the peak frequencies sufficiently different.\\

However, the amplitude of these scenarios is smaller than that of the standard scenario for equal maximum temperature, as discussed before. 
Heuristically, this setup achieves to obtain a CGWB dominated by the hidden sector by diminishing the contribution from the visible sector.
Since the SM prediction is still quite far from the reach of proposed experiments~\cite{Aggarwal:2020olq}, we do not study this scenario any further. 
The truly new and interesting results appear when considering an exotic equation of state, to which we now turn.

\subsection{Scenario 2 -- Stiff equation of state}\label{sec:kination}

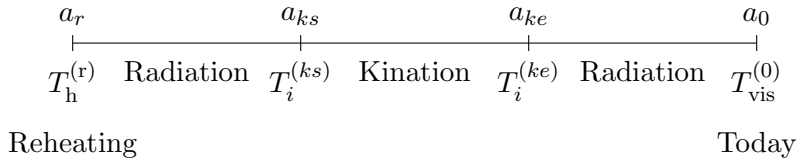
\begin{figure}[t!]
\centering
  \begin{tikzpicture}[]

    \draw (0,0) -- (9,0);

    \foreach \x in {0,3,6,9}
      \draw (\x cm,3pt) -- (\x cm,-3pt);

    \draw (0,0) node[below=3pt] {$ \tem{\text{h}}{\text{r}} $} node[above=3pt] {$a_r$};
    \draw (3,0) node[below=3pt] {$ \tem{i}{ks} $} node[above=3pt] {$a_{ks}$};
    \draw (6,0) node[below=3pt] {$ \tem{i}{ke} $} node[above=3pt] {$a_{ke}$};
    \draw (9,0) node[below=3pt] {$ \tem{\text{vis}}{0} $} node[above=3pt] {$a_0$};
    \draw (0,-1) node[below=1pt] {Reheating};
    \draw (9,-1) node[below=1pt] {Today};
    \draw (1.5,0) node[below=3pt] {Radiation};
    \draw (4.5,0) node[below=3pt] {Kination};
    \draw (7.5,0) node[below=3pt] {Radiation};
  \end{tikzpicture}
  \caption{Timeline of the evolution of a sector undergoing kination. The dominant contribution to the CGWB is sourced during the first radiation period, right after reheating, where we integrate Eq.~\eqref{eq:ampl} up to the time given by $a_{ks}$. \label{fig:timeline}}
\end{figure}

In this subsection we study how the shape of the CGWB is modified if the energy density of the hidden sector is `cooled down'. We model this by a sudden change in its equation of state $P=\omega \rho$, $\omega>1/3$, yielding
\begin{equation}\label{eq: rho redshift}
    \rho(t)=\rho_1\lr{\frac{a_1}{a(t)}}^{3(1+\omega)} \,.
\end{equation}
The behaviour of undergoing such a change is sketched in the blue solid line in Fig.~\ref{fig:scenario3}, in the example of a period of \textit{kination} \cite{Gouttenoire:2021jhk}, with equation of state $\omega=1$, which is the maximum value allowed by causality. This scenario is well motivated as a period between inflation and reheating, and in models with axions, where kination follows a matter-dominated era.
The former is not much of our interest since the CGWB is emitted after reheating.
We henceforth restrict ourselves to a cosmological history of the form radiation-kination-radiation, as depicted in Fig.~\ref{fig:timeline},
though our results could be easily generalised to more advanced settings. For recent interesting work on kination inspired by string cosmology see \cite{Conlon:2022pnx, Apers:2022cyl}, and~\cite{Chowdhury:2022gdc} for the study of a kination-like period motivated in scalar-tensor theories, and its consequences the spectrum of primordial gravitational waves.

This scenario strongly alters the dynamics of spacetime
since the expansion of the Universe is slowed down in a period of kination domination $H^2\sim a^{-6}$.
However, the causal structure of FLRW is preserved and the GWs follow the same geodesics, redshifting as $\rho \sim a^{-4}$. The shape of $\Omega_{\tti{GW},0} (f)$ emitted by the hottest sector is thus unaltered and Eq.~\eqref{eq: omega f} still applies.\footnote{If this signal has been emitted by a hidden sector, the signals emitted during kination domination from other sectors will be altered because entropy is not conserved, so we cannot use Eq.~\eqref{eq:OmegaSector}. These are, however, subdominant with respect to the former.}\\

To quantify the effects of the kination epoch,
we focus on the GW spectrum $\Omega_{\tti{GW},0}^{(i)}$ of the hottest sector and compare it against the spectrum $\tilde{\Omega}_{\tti{GW},0}^{(i)}$ of the equivalent sector in a putative situation without kination. 

One then finds from Eq.~\eqref{eq:ampl}
\begin{equation}
    \frac{ \Omega_{\tti{GW},0}^{(i)}}{ \tilde{\Omega}_{\tti{GW},0}^{(i)}} \sim \lr{\frac{a_r \tem{i}{r}} {\tilde{a}_{r}\tilde{T}_i^{(r)}}}^4 \,.
\end{equation}
For simplicity,
we assume that the sectors have the same energy density at BBN which implies
\begin{equation}
    \tilde{a}_r \tilde{T}_i^{(r)}=a_{ke}\tem{i}{ke} \, ,\quad a_r T_i^{(r)} = a_{ks} \tem{i}{ks}  \, ,
\end{equation}
where $a_{ks}$ ($a_{ke}$) stands for the start (end) of kination.
We can then
trace the evolution of $T$ from \eqref{eq: rho redshift} by assuming that the changes are instantaneous, thereby giving rise to
\begin{equation}
    \frac{\tem{i}{ks}}{\tem{i}{ke}}=\lr{\frac{a_{ke}}{a_{ks}}}^{3(1+\omega)/4}\, .
\end{equation}
Hence, the ratio of the amplitudes only depends on the duration of the kination period
\begin{equation}\label{eq: omega kination}
    \frac{ \Omega_{\tti{GW},0}^{(i)}}{ \tilde{\Omega}_{\tti{GW},0}^{(i)}} \sim \lr{\frac{a_{ke}}{a_{ks}}}^{3\omega-1}\, .
\end{equation}
In words, this suggests that a period with equation of state $\omega>1/3$ significantly boosts the GW spectrum.
For a given sector and $\omega$, the resulting parameter space containing the number of e-folds and initial temperature is then constrained by BBN bounds analogously to Eq.~\eqref{eq:BBNBound}.
A similar analysis for the frequencies gives 
\begin{equation}
    \dfrac{f}{\Tilde{f}}=\lr{\frac{a_{ke}}{a_{ks}}}^{\frac{3\omega-1}{4}} \,,
\end{equation}
implying that the spectrum is blue-shifted compared to the standard scenario.\\

\begin{figure}[t!]
\centerline{\includegraphics[width=0.7\textwidth]{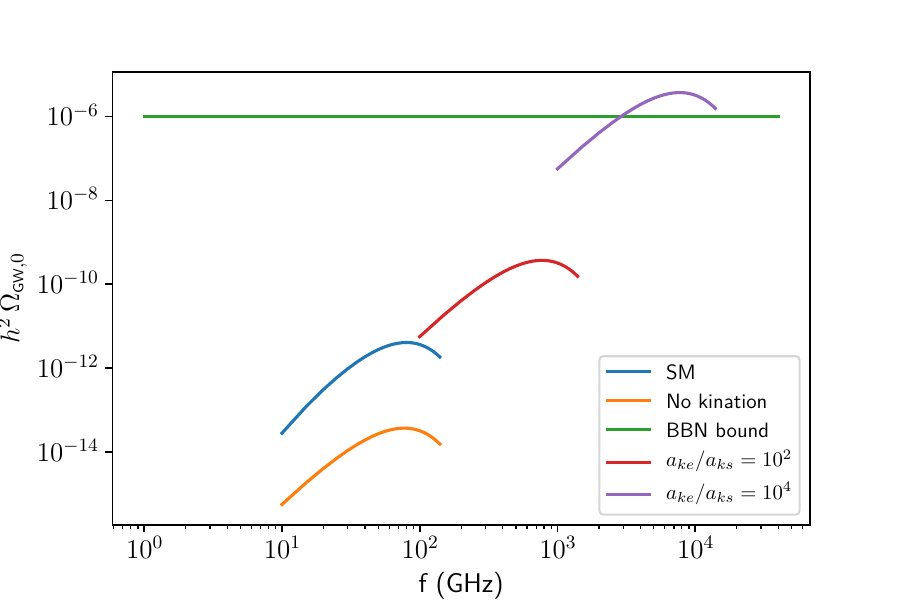}}
\caption{$h^2 \Omega_{\tti{GW},0}$ as a function of frequency in the kination scenario. The plot shows the GW spectrum of a kinating, hidden $\text{SU}(2)$ for different e-folds of kination. If kination is too long, the BBN bound is violated. The standard scenario with only the SM is illustrated for comparison. }
\label{fig: kination and bounds}
\end{figure}

Let us consider a hidden sector described by pure Yang-Mills $\text{SU}(2)$, with hottest temperature $T=10^{-4}M_P$.
If there was no kination period in the early Universe, the CGWB sourced by this sector is negligible\footnote{And the BBN bound is satisfied for the SM and the hidden sector heated up at the same temperature, since $\rho_h/\rho_{\tti{BBN}}=24/(7\Delta N_{\tti{eff}})\lr{g_{*,\tti{BBN}}/g_*}^{4/3}\sim0.53$.}. 
If such a period happened, however, the GW signal emitted by the self-interactions can even saturate the BBN bound, as observed in Fig.~\ref{fig: kination and bounds}.
We thus find that there is a maximum number of kination e-folds that we can allow for, given a maximum temperature. 
In this case, $\tem{h}{1}=10^{-4}M_{P}$, we find that a kination period longer than about $8.5$ e-folds $\lr{a_{ke}/a_{ks}\sim 10^{3.7}}$ is ruled out. 

We can give precise bounds using the knowledge of the peak of $h^2\Omega_{\tti{GW},0}$ in the standard scenario with a given maximum temperature $\tilde{T}$. Let us denote this peak by $h^2\tilde{\Omega}_{\tti{GW},0}$.

\begin{figure}[t!]
\centerline{\includegraphics[width=0.7\textwidth]{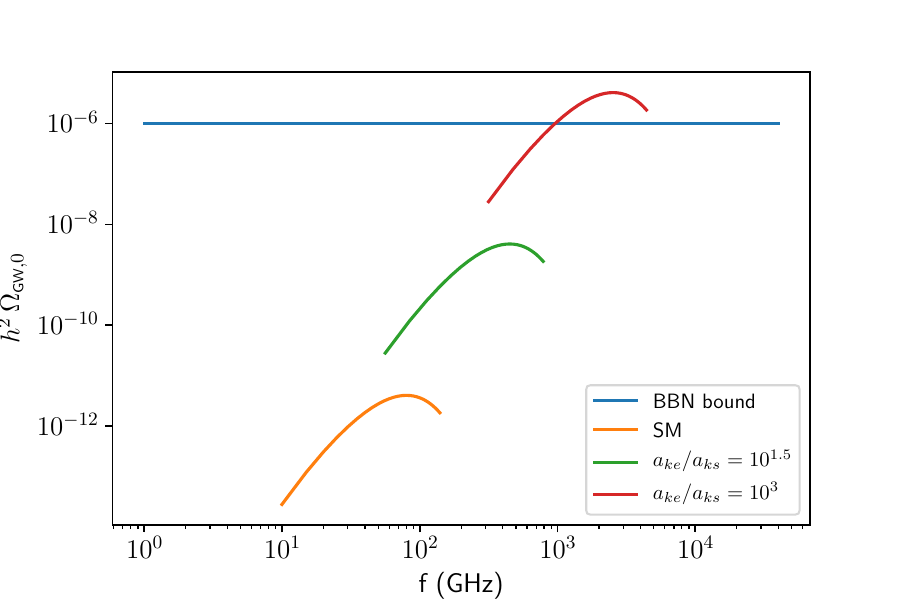}}
\caption{$h^2 \Omega_{\tti{GW},0}$ as a function of frequency in the kination scenario, in the presence of only one sector, given by the SM. The number of e-folds of kination is more constrained in this case.}
\label{fig:smkination}
\end{figure}

Because the CGWB is linear with the maximum temperature, Eq.~\eqref{eq: omega kination} generalises to
\begin{equation}\label{eq: omega kination 2}
    \frac{ \Omega_{\tti{GW},0}}{ \tilde{\Omega}_{\tti{GW},0}} \sim \frac{T}{\Tilde{T}}\lr{\frac{a_{ke}}{a_{ks}}}^{3\omega-1} \, ,
\end{equation}
where for ease of notation $T$ denotes the reheating temperature in the kination scenario, and we have neglected subdominant contributions to the amplitude from the other sectors. 
In our example, $h^2\tilde{\Omega}_{\tti{GW},0}\sim 10^{-14}$
for $\Tilde{T}=10^{-4}M_{P}$. 
Imposing that $h^2 \Omega_{\tti{GW},0}<10^{-6}$ in the kination scenario, we obtain a simple bound in the ratio of scale factors:
\begin{equation}
    \frac{T}{M_{P}}\lr{\frac{a_{ke}}{a_{ks}}}^{3\omega-1}< \frac{10^{-6}}{h^2\tilde{\Omega}_{\tti{GW},0}}\frac{\tilde T}{M_{P}}\sim10^4 \, .
\end{equation}
Another example would be the SM undergoing kination. 
Such a scenario is harder to motivate, because of the complexity of the field content of the SM. 
Indeed, kination is usually considered for scalar fields, and it can simply understood as the field quickly rolling down its potential.
We can motivate kination for gauge fields\footnote{This is the most interesting case as it is the gauge fields that source the dominant, HTL contribution to the GW spectrum.} if there is an equivalent description of them in terms of a scalar field, such as after confinement. 
Thus, the previous example could be understood as a weakly coupled YM theory that undergoes confinement, and the effective scalar field undergoes kination.
We cannot motivate a similar setup where the whole field content of the SM is described by a scalar field, or that it undergoes a kination period otherwise.
However, it was pointed out in~\cite{Chowdhury:2022gdc} that an effective period of kination can appear in scalar-tensor theories. 
An analysis for the consequences of the CGWB in such scalar-tensor theories is beyond the scope of this paper, but we believe that our considerations can yield an interesting phenomenology of such scenarios.
We thus provide an example in which the SM undergoes a kination-like period in Fig.~\ref{fig:smkination}. 
For the a maximum temperature $T=10^{-4}M_P$, we find that more than 6 e-folds of kination $\lr{a_{ke}/a_{ks}\sim 10^{2.7}}$ would be ruled out.

\section{Conclusions}\label{sec:conclusions}

We conclude that the CGWB has the potential to test BSM physics, carrying an imprint of the first hot radiation content of the Universe. Being an irreducible background, its detection could  be thought of as a milestone to future detectors, albeit most probably in the very  long term.  
It is worth reiterating that these signals were sourced in the very early Universe, right after reheating, by a radiation bath in thermal equilibrium. Potential high-resolution detectors (and calculations) could be used to study the interactions of hidden sectors through gravity. The CGWB thus sets yet another case \cite{Aggarwal:2020olq} of studying particle physics with GWs. 
We claim that a measurement 
\begin{itemize}
\item around the SM characteristic frequency $f \sim 80$ GHz would imply a strong case for the absence of new physics after reheating, and determine the reheating temperature. 
\item at a smaller frequency would imply either the existence of many massive degrees of freedom, or potentially the existence of a hidden sector. 
\item at a higher frequency would imply a period with a stiff equation of state.
\end{itemize}

Our analysis also rules out a parameter space of e-folds and maximum temperature in models of the form radiation-kination-radiation. These constraints appear for models in which the radiation bath includes gauge fields, since we only considered the hard thermal loop approximation. We point out, however, that similar considerations can be applied to any model with a stiff equation of state that sources GWs. These issues certainly deserve further investigations.

\section*{Acknowledgements}

We would like to thank Debika Chowdhury, Andreas Ringwald, Carlos Tamarit, Gianmassimo Tasinato and Ivonne Zavala for useful discussions.
AS thanks DAMTP at the University of Cambridge for hospitality where parts of this work have been completed. 
The research of AS was in part supported by NSF grant PHY-2014071.
FM is funded by a UKRI/EPSRC Stephen Hawking fellowship, grant reference EP/T017279/1, partially supported by the STFC consolidated grant ST/P000681/1 and funded by a G-Research grant for postdocs in quantitative fields.
 The work of FQ has been partially supported by STFC consolidated grants ST/P000681/1, ST/T000694/1. The work of GV has been partially supported by STFC consolidated grant ST/T000694/1.

\appendix 

\section{Running through stages}\label{sec: integrate omega}

This Appendix is devoted to collect the relevant formulae for a higher-precision calculation. The source term $\heta$ receives three contributions from gauge, matter and HTL effects.
The former give rise to
\begin{align}\label{eq:eta_g}
\heta_{\text{\tiny{g}}}(T, \hk) =  \sum_{n=1}^{N_g} g^2_n(T) \, \mathcal{N}_n \bigg\lbrace \frac{1}{2} T_{n,Ad} \, \eta_{gg}(\hk) + \sum_{p}{T_{n,p} \, \eta_{sg}(\hk) + \frac{1}{2} \sum_{\gamma}{T_{n,\gamma} \, \eta_{fg}(\hk)}} \bigg\rbrace \,,
\end{align}
where we consider a theory with $n=1,\ldots,N_g$ gauge groups with $\mathcal{N}_n$-dimensional Lie algebras.
For every gauge field there are Weyl fermions $\psi_{\gamma}$, $\gamma=1,\ldots ,N_{f}$, transforming under a representation with Dynkin index\footnote{The Dynkin index is defined as $\Tr(\hat{T}(R)^a \hat{T}(R)^b)=T_R\delta^{ab}$, where $\hat{T}(R)^a$ is the $a^{th}$ generator of the representation $R$ of the Lie algebra of the relevant group. For instance, in $\text{SU}(N_c)$ the Dynkin index equals $1/2$ in the fundamental and $N_c$ in the adjoint representation.} $T_{n,\gamma}$, and similarly scalars $\phi_{p}$, $p=1,\ldots ,N_{s}$, with Dynkin indices $T_{n,p}$. 
The functions $\eta_{AB}$ parametrise interactions among particle species $A$ and $B$ up to two loops, see App.~A of~\cite{Ringwald:2020ist} for explicit expressions. 

The matter part of $\heta$ reads
\begin{equation}\label{eq:eta_m}
\heta_{\tiny{m}}(T, \hk) =  \frac{1}{2}\sum_{p, \gamma, \delta} {|y^p_{\gamma\delta}(T)|^2\, \heta_{sf}(\hk)}
\end{equation}
and is fully specified by Yukawa couplings $ y^p_{\gamma\delta}$.

Lastly, the expression for the thermal masses appearing in the HTL part, is given by
\begin{equation}
    m_n^2 (T) =T^{2}\hat{m}_n^2(T) \,,\qquad \hat{m}_n^2(T)=  g_n^2(T)  \left \lbrace \frac{1}{3}T_{n,Ad} + \frac{1}{6}\sum_{p}{T_{n,p}}+\frac{1}{6}\sum_{\delta}{T_{n,\delta}} \right \rbrace\,.
\end{equation}
The temperature dependence of $m_n(T)$ implies that they are relevant during the whole expansion of the Universe.
In contrast,
bare masses of the fields are only relevant at particular temperatures, namely when the associated thermal energy is comparable to such mass.

On this point, note that the process of fields becoming massive necessarily modifies the effective degrees of freedom, so in this regime our approximations fail and the integral in Eq.~\eqref{eq:ampl} is not immediate.
We can perform an analysis of the contributions throughout cosmic history by restricting the analysis to $N_s$ \textit{stages} in the history of the Universe in which the number of relativistic degrees of freedom $g_{*,i}$ in the $i$-th sector is constant.

In practice,
the integral in Eq.~\eqref{eq:ampl} splits into a sum over stages labelled by $\alpha$
\begin{equation}\label{eq:stages}
    \Omega^{(i)}_{\text{\tiny{GW}},0}= \sum_{\alpha=1}^{N_s}\Omega^{(\alpha,i)}_{\text{\tiny{GW}},0} + \tilde\Omega^{(i)}_{\text{\tiny{GW}},0} \,,
\end{equation}
where the term $\tilde\Omega^{(i)}_{\text{\tiny GW},0}$ accounts for contributions from non-constant $g_{*,i}$. Explicitly, the terms involved are:
\begin{align}
h^{2}\Omega^{(\alpha,i)}_{\text{\tiny{GW}},0} &= \frac{\lambda}{M_{P}} \frac{h^2 \Omega_{\gamma, 0}}{(a_0 T_{{\rm vis},0})^4} 
    \int_{t^{\alpha-1}+\varepsilon}^{t^{\alpha}-\varepsilon}{\dif a \, \frac{(aT_i)^5}{a^2} \, \dfrac{T_{i}^{2}}{\sqrt{\rho}}\, \hk_i^3 \heta_i (T_i, \hk_i ) } \,,\\[0.5em]
h^{2}\tilde\Omega^{(i)}_{\text{\tiny{GW}},0} &= \frac{\lambda}{ M_{P}}  \frac{h^2 \Omega_{\gamma,0}}{(a_0 T_{{\rm vis},0})^4}
    \sum_{\alpha}\int_{t^{\alpha}-\varepsilon}^{t^{\alpha}+\varepsilon}{\dif a \, \frac{(aT_i)^5}{a^2} \, \dfrac{T_{i}^{2}}{\sqrt{\rho}}\, \hk_i^3 \heta_i (T_i, \hk_i )  } \,,
\end{align}
where $(t^{\alpha}-\varepsilon,t^{\alpha}+\varepsilon)$ indicates the time interval in which the transition between the $\alpha^{th}$ and $(\alpha+1)^{th}$ stages occurs. In this time interval, not all the relativistic degrees of freedom $g_{*,i}$ are constant, and so the integration is not immediate. A numerical improvement could be achieved performing a Daisy resummation as in \cite{Ringwald:2020vei}. 

Furthermore, since the $\Omega^{(\alpha,i)}_{\tti{GW,0}}$ are linear with the maximum temperature of the stage, it follows that the maximum contribution to the CGWB comes from the first stage after reheating, in which we have focused on the main text. 

\bibliographystyle{utphys}
\bibliography{biblio}

\end{document}